\documentclass[%
 reprint,
 amsmath,amssymb,
 aps,
longbibliography,
]{revtex4-1}

\usepackage{upgreek}
\usepackage{graphicx}
\usepackage{dcolumn}
\usepackage{bm}
\usepackage{color}
\usepackage{mathtools}
\usepackage[normalem]{ulem}
\usepackage[english]{babel}
\usepackage{xr-hyper}
\usepackage[colorlinks=true,linkcolor=blue,urlcolor=blue,citecolor=blue]{hyperref}
\usepackage[toc,page]{appendix}

\let\eq\relax

\begin{document}

\title{Large heat-capacity jump in cooling-heating of fragile glass from kinetic Monte Carlo simulations based on a two-state picture}

\author{Chun-Shing Lee$^1$}
\author{Hai-Yao Deng$^2$}
\author{Cho-Tung Yip$^3$}
\author{Chi-Hang Lam$^1$}
\email[Email: ]{C.H.Lam@polyu.edu.hk}
\address{
	$^1$Department of Applied Physics, Hong Kong Polytechnic University, Hong Kong, China \\
	$^2$School of Physics and Astronomy, Cardiff University, 5 The Parade, Cardiff CF24 3AA, Wales, UK\\
    $^3$Department of Physics, Shenzhen Graduate School, Harbin Institute of Technology, Shenzhen 518055, China 
  }

\date{\today}

\newcommand{\blue}[1]{{\color{blue}{#1}}}
\newcommand\cmt[1]{\textcolor{red}{[#1]}}
\newcommand\move[1]{\textcolor{blue}{[#1]}}
\newcommand{\red}[1]{{\color{red}{#1}\color{black}}}
\newcommand{\chg}[1]{{\color{blue}{#1}}}        
\newcommand{\del}[1]{\sout{#1}}  
\renewcommand{\del}[1]{}  
\newcommand{\delete}[1]{\sout{#1}}  
\newcommand{\rpl}[2]{{\sout{#1}}{\color{blue}{#2}}}        
\newcommand{\dis}[1]{{\color{red}\del{#1}}} 
\newcommand{\chk}[1]{{\color{red}\{? #1\}}}
\newcommand{\qn}[1]{{\color{red}\{#1 \}}}

\newcommand{\refb}[1]{(\ref{#1})}
\renewcommand{\Ref}[1]{Ref.~\cite{#1}}
\newcommand{\fig}[1]{Fig.~\ref{#1}}
\newcommand{\figs}[2]{Figs.~\ref{#1} and \ref{#2}}
\newcommand{\Fig}[1]{Figure~\ref{#1}}
\newcommand{\eq}[1]{Eq.~(\ref{#1})}
\newcommand{\eqr}[2]{Eqs.~(\ref{#1})-(\ref{#2})}
\newcommand{\Eqr}[2]{Equations~(\ref{#1})-(\ref{#2})}
\newcommand{\eqs}[2]{Eqs.~(\ref{#1}) and (\ref{#2})}
\newcommand{\Eqs}[2]{Equations~(\ref{#1})-(\ref{#2})}
\newcommand{\Eq}[1]{Equation~(\ref{#1})}
\newcommand{\App}[1]{Appendix~\ref{#1}}
\renewcommand{\sec}[1]{Sec.~\ref{#1}}
\newcommand{\Sec}[1]{Section~\ref{#1}}
\newcommand{\heading}[1]{~\\\noindent{\bf #1:}}

\newcommand{\roundbk}[1]{\left({#1}\right)}
\newcommand{\squarebk}[1]{\left[{#1}\right]}
\newcommand{\bracebk}[1]{\left\lbrace{#1}\right\rbrace}
\newcommand{\anglebk}[1]{\left\langle{#1}\right\rangle}
\newcommand{\lbracebk}[1]{\left\lbrace{#1}\right.}
\newcommand{\rbracebk}[1]{\left.{#1}\right\rbrace}

\newcommand{\phiv}{\phi_{v}}
\newcommand{\MSD}{\anglebk{|\mathbf{r}_{l}(t) - \mathbf{r}_{l}(0)|^{2}}}
\newcommand{\dE}{\Delta E}
\newcommand{\DV}{\Delta V}
\newcommand{\mkmin}{m_{k}^{strong}}
\newcommand{\mkmax}{m_{k}^{fragile}}

\newcommand{\Dref}{10^{-1}}

\begin{abstract}
The specific heat capacity $c_v$ of glass formers undergoes a hysteresis when subjected to a cooling-heating cycle, with a larger $c_v$ and a more pronounced hysteresis for fragile glasses than for strong ones.
Here, we show that these experimental features, including the unusually large magnitude of $c_v$ of fragile glasses, are well reproduced by kinetic Monte Carlo and equilibrium study of a distinguishable particle lattice model (DPLM) incorporating a two-state picture of particle interactions. The large $c_v$ in fragile glasses is caused by a dramatic transfer of probabilistic weight from high-energy particle interactions to low-energy ones as temperature decreases.
\end{abstract}

\maketitle
\section{Introduction}
Many fascinating aspects of glass transition rest with their non-equilibrium nature as seen in the history dependence of the thermodynamic and kinetic behaviors of glass formers \cite{biroli2013review,stillinger2013review}.
In this work, we study long known puzzles related to their specific heat capacity $c_v$.
When subjected to a cooling-heating cycle, $c_v$ exhibits a rather abrupt jump between corresponding values for liquid and glass close to the glass transition temperature.
Glasses can be broadly classified as fragile or strong, depending on the degree of deviation from Arrhenius behaviors.
Perplexingly, the jump magnitude of $c_v$ is surprisingly large for fragile glasses, such as typical organic and polymeric glasses, and can reach a few $k_B$, where $k_B$ is the Boltzmann constant. It is in contrast much smaller for strong glasses such as silicates \cite{angell2011}.  
In addition, one also observes hysteresis in $c_v$ during heating and cooling \cite{moynihan1974,hodge1994,keys2013,li2017b,zheng2019,chen2009,tropin2018book}, which is much more pronounced for fragile glasses \cite{li2017b,tropin2018book}.
Phenomenological descriptions of $c_v$ and the hysteresis have been advanced by mean-field theories based on a fictive temperature \cite{hodge1994,tanaka2017}.
A fundamental reason for the dependence on fragility remains elusive. 
The phenomena have so far lacked atomistic simulations. One challenge, for example, is that   
molecular dynamics (MD) simulations \cite{kremer1990,kob1995} can hardly cope with sufficiently low cooling/heating rates entailing long computational time.

Lattice models play pivotal roles in many branches of statistical physics as they highlight the essential physics and achieve superior computational speed via omitting irrelevant details \cite{redner2010,binderbook}. To study glasses, most lattice models, including kinetically constrained models (KCM) \cite{fredrickson1984,palmer1984} and lattice glass models \cite{biroli2001}, focus primarily on the kinetics and are energetically trivial with a vanishing $c_v$ \cite{ritort2003review,garrahan2011review}. Generalizations to energetic variants however give 
$c_v$ way smaller than typical values observed for fragile glasses \cite{fredrickson1986,mccullagh2005,nishikawa2020}. It was argued that defect models, as most of these models are, cannot intrinsically capture the thermodynamics of glass \cite{biroli2005}. For example, $c_v$ in the two-spin facilitation model, an important KCM,  is proportional to the defect density and thus becomes very small at low temperatures \cite{fredrickson1986}. Only after  coupling a lattice model to a fictive temperature field in an \textit{ad hoc} fashion,   $c_v$ can be freely fine-tuned and match realistic values  \cite{keys2013}.
Nevertheless, it appears that lattice models by themselves, without any coupled field, are intrinsically incapable of capturing the correct magnitude of $c_v$. 
This severely limits their usefulness in studying glass thermodynamics and, strangely, is at odds with the stronger roles of lattice models in many other branches of statistical physics  \cite{redner2010,binderbook}. In addition, conventional lattice models in general cover only a limited range of fragility. This imposes another major difficulty in investigating the fragility-dependence of glass thermodynamics.

Here, we show that a recently proposed distinguishable particle lattice model (DPLM) \cite{zhang2017} naturally captures the major experimentally observed thermodynamic features of glass, including the large value of $c_v$ and the hysteresis. The close correlation of these features with fragility is also clearly demonstrated.
This is made possible by the capability of the DPLM to simulate both strong and fragile glasses, with respective characteristic properties already demonstrated to be consistent with experimental trends \cite{lee2020}.  

\section{Model}
The DPLM assumes $N$ hard-core particles, each representing a rigid molecular group of atoms. They live on a square lattice of size $L^2> N$ with the lattice constant set to unity.
Each particle is of its own species indexed by $s = 1,2,...,N$ and hence distinguishable. 
The total energy of the system is given by
\begin{equation}
	E = \sum_{<i,j>'} V_{s_i s_j},
	\label{E}
\end{equation}
where the sum runs over all pairs of nearest neighboring sites occupied by particles. The energy per particle is $\varepsilon = E/N$.  

Thermodynamic properties of glasses of various fragilities have been successfully accounted for using a simple two-state model, also called the bond-excitation model, of Moynihan and Angell, which assumes particle interactions taking independently one of two possible strengths \cite{moynihan2000}. 
To incorporate the two-state picture into the fully atomistic and dynamical DPLM, we sample the particle interaction energies $V_{kl}$ for all particle pairs $k$ and $l$ from a \textit{bi-component}  form of the interaction distribution $g(V)$ before a simulation commences.
It consists of a uniform low-energy part and a sharp high-energy component represented by a delta function,
\begin{equation}
	g(V) = \frac{G_0}{\DV} + (1-G_0) \delta(V-V_1),
	\label{gV}
\end{equation}
where $V\in[V_0,V_1]$ with $V_0 = -V_1 = - 0.5$ and $\DV = V_1-V_0 = 1$
serves as the unit of energy. 
In addition, $\delta$ denotes the Dirac function which may be replaced by some narrowly peaked distributions (e.g. a Gaussian) without affecting the results, and $G_0 \in [0,1]$ is an energetic parameter that controls the thermodynamic properties of the system.
It has been shown that a smaller (but finite) $G_0$ leads to fragile glasses while a bigger $G_0$ to strong glasses. By tuning $G_0$, a wide range of fragility can be realized \cite{lee2020}.

Unlike many other lattice models, the DPLM is intrinsically a particle model, a property essential for the direct study of the thermodynamics as particles, rather than defects, should dominate the system energy.
A defect in the DPLM is instead represented implicitly by the absence of a particle, i.e. a void. 
We envision a void as a unit of free volume, which was long known to be important in glassy dynamics \cite{cohen1961}. Its relevance has been disputed more than a decade ago \cite{harrowell2006}. However, sophisticated machine learning approaches have recently correlated mobility with local particle density \cite{ma2019,bapst2020}. We have also directly identified quasi-voids, each consisting of localized and fragmented free volumes, in colloidal experiments at very high packing fractions \cite{yip2020}.
Using the Metropolis algorithm, a particle can move to an adjacent void at the following rate \cite{lee2020}
\begin{equation}
	w = 
	\begin{cases}
		w_0\exp(-\dE/k_BT), & \mbox{for}~\dE > 0 \\
		w_0, & \mbox{for}~\dE \leq 0
	\end{cases}
	\label{w}
\end{equation}
where $\dE$ is the change in the system energy due to the hop, $T$ is the bath temperature, $w_0 = 10^6$ is the attempt frequency and $k_B=1$ is the Boltzmann constant.

\begin{figure}[t]
	\includegraphics[width=\linewidth]{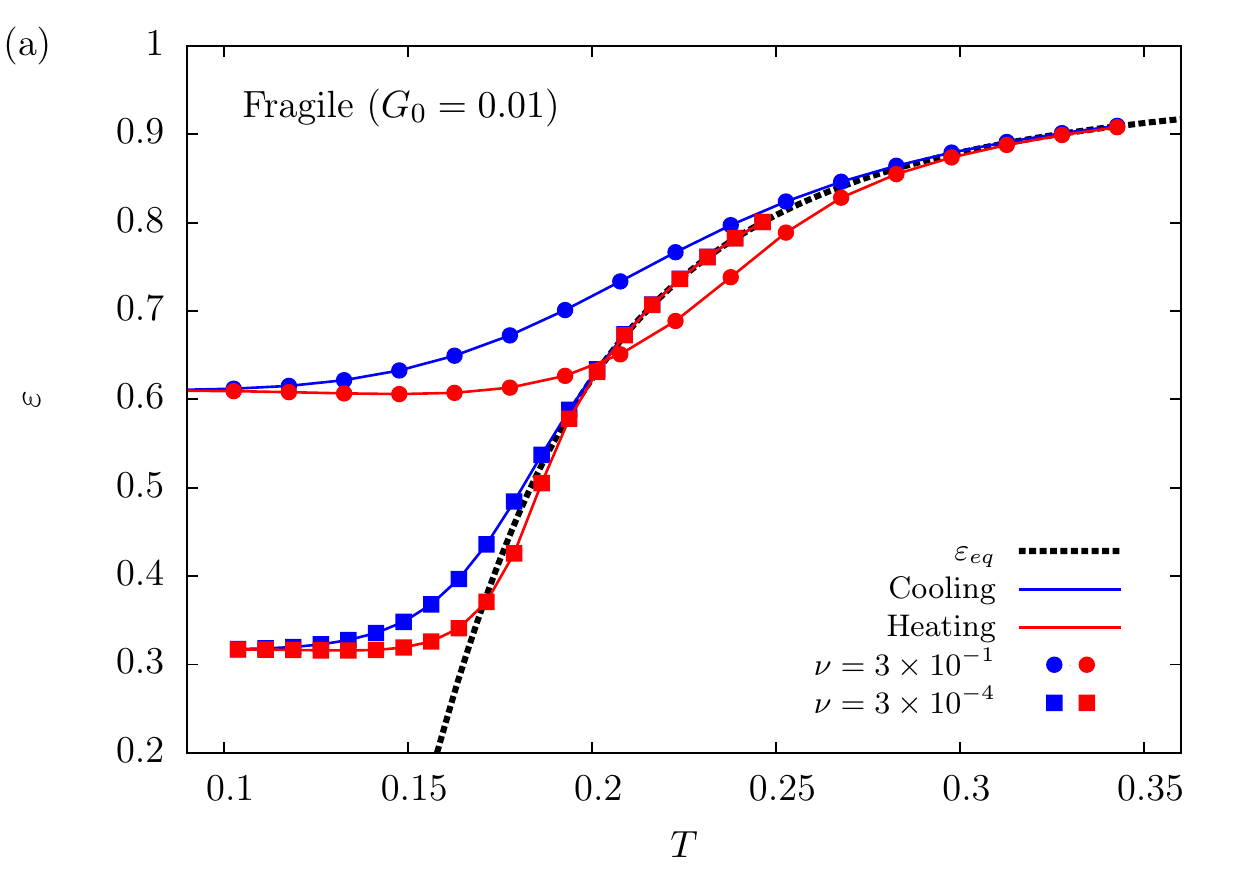}
	\includegraphics[width=\linewidth]{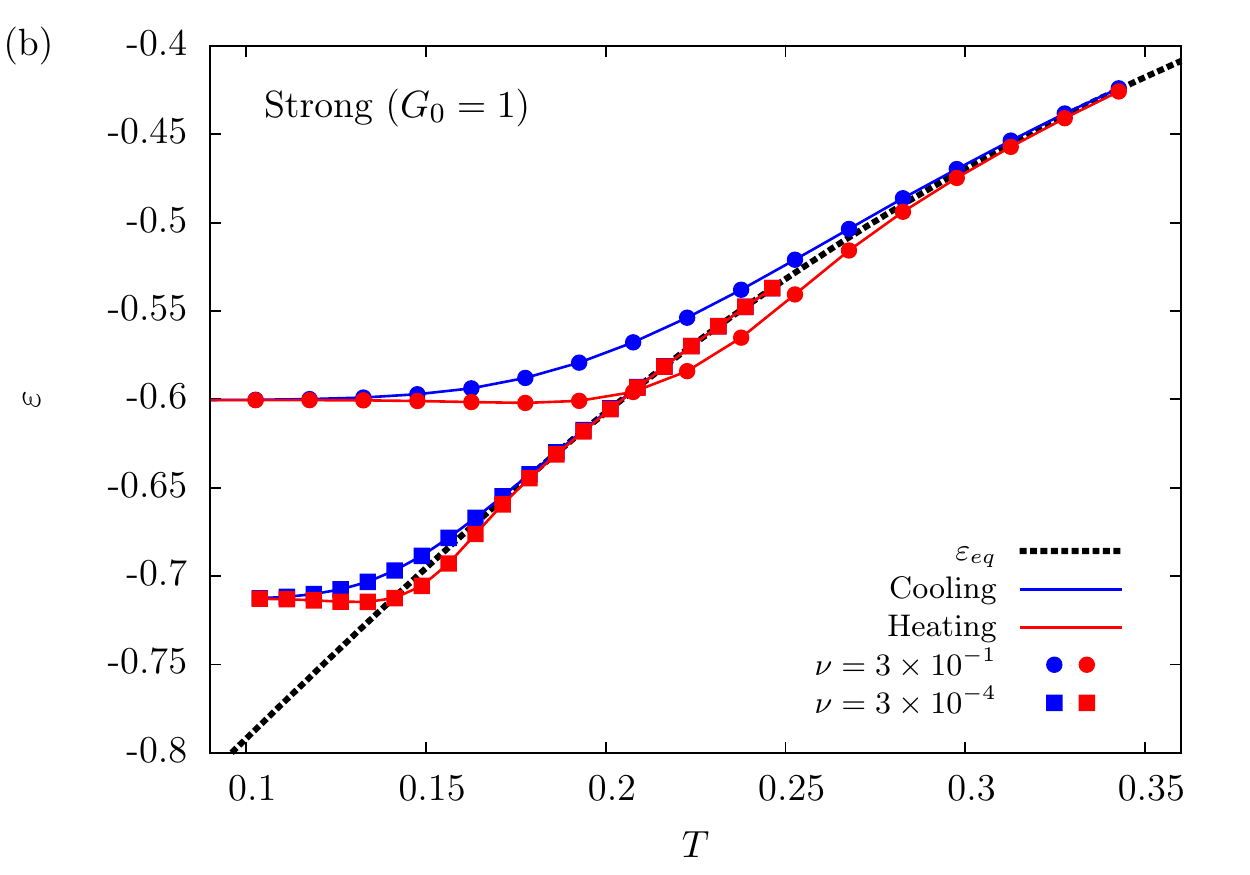}
	\caption{
		Energy per particle $\varepsilon$ during a cooling-heating cycle for (a) fragile glass with $G_0 = 0.01$ and (b) strong glass with $G_0 = 1$ at cooling/heating rate $\nu$. 
		The black dashed lines show equilibrium energy $\varepsilon_{eq}$ calculated from \eq{Eeq}.
	}
	\label{fig:eT}
\end{figure}

\section{Calorimetric analysis}
We focus mainly on $G_0 = 0.01$ and 1. As indicated by an Angell plot and by  extrapolating simulation results to realistic time scales, they are found to have kinetic fragility of about 116 and 31 respectively, modeling fragile and moderately strong glasses (see Appendix \ref{appendix:glassy_dynamics}).
Very strong glasses can also be modeled by introducing an additive offset to the particle hopping energy barrier or by using a different form of $g(V)$ and will be studied in the future.
%
We subject the system to a cooling-heating cycle and study its out-of-equilibrium calorimetric responses.
Using a direct construction method \cite{zhang2017}, we first prepare the system in thermodynamic equilibrium at some temperature $T_0$  much higher than the glass transition temperature $T_g$. 
Then, we lower the bath temperature $T$ at a constant rate
$\nu = |d T/d t|$.
Once $T$ decreases to a temperature much lower than $T_g$, we reverse the process and heat up the system at the same rate $\nu$ until $T$ reaches $T_0$.

\subsection{Energy and heat capacity}
The energy per particle $\varepsilon$ is monitored throughout the entire cooling-heating cycle. 
The specific heat $c_v$ at constant volume is then calculated from $c_v = d \varepsilon/d T$.

\begin{figure}
	\includegraphics[width=\linewidth]{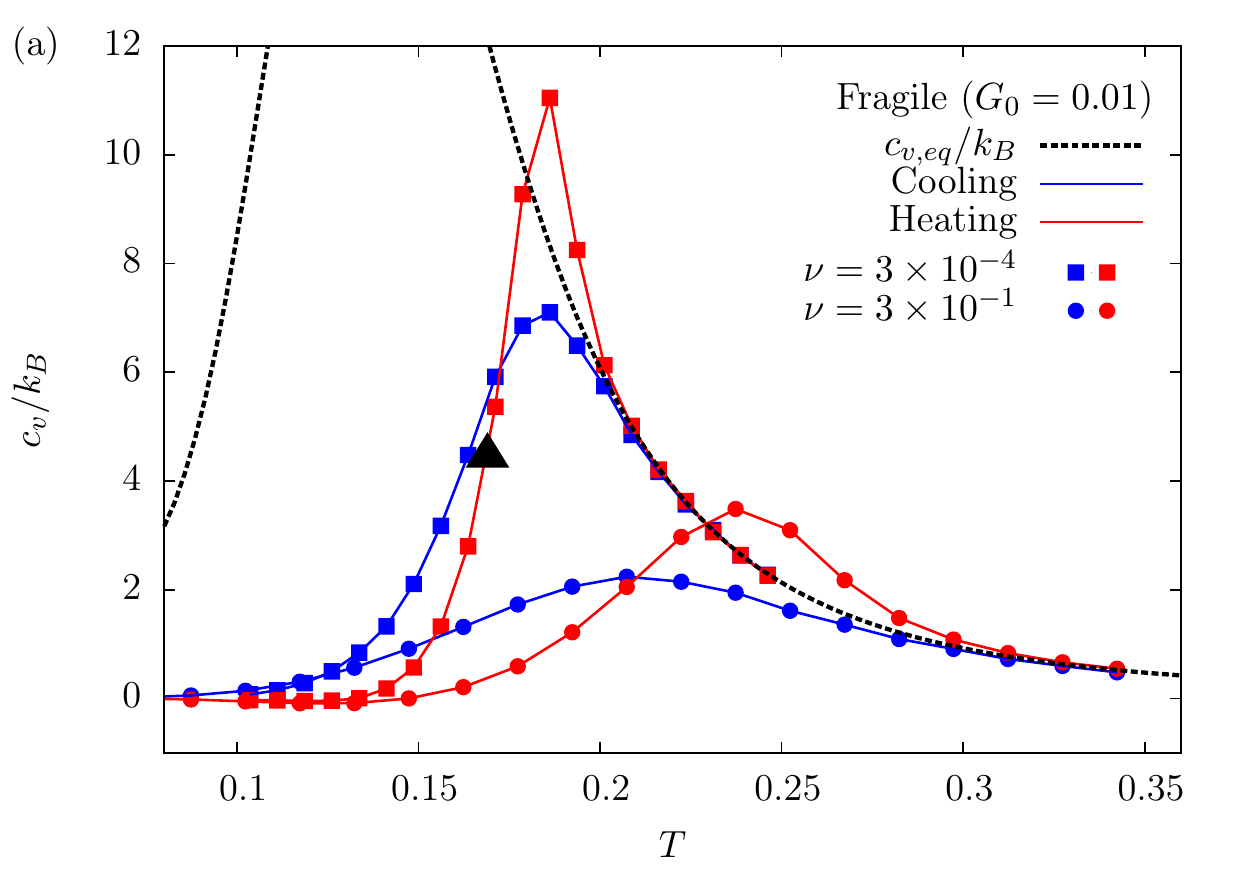}
	\includegraphics[width=\linewidth]{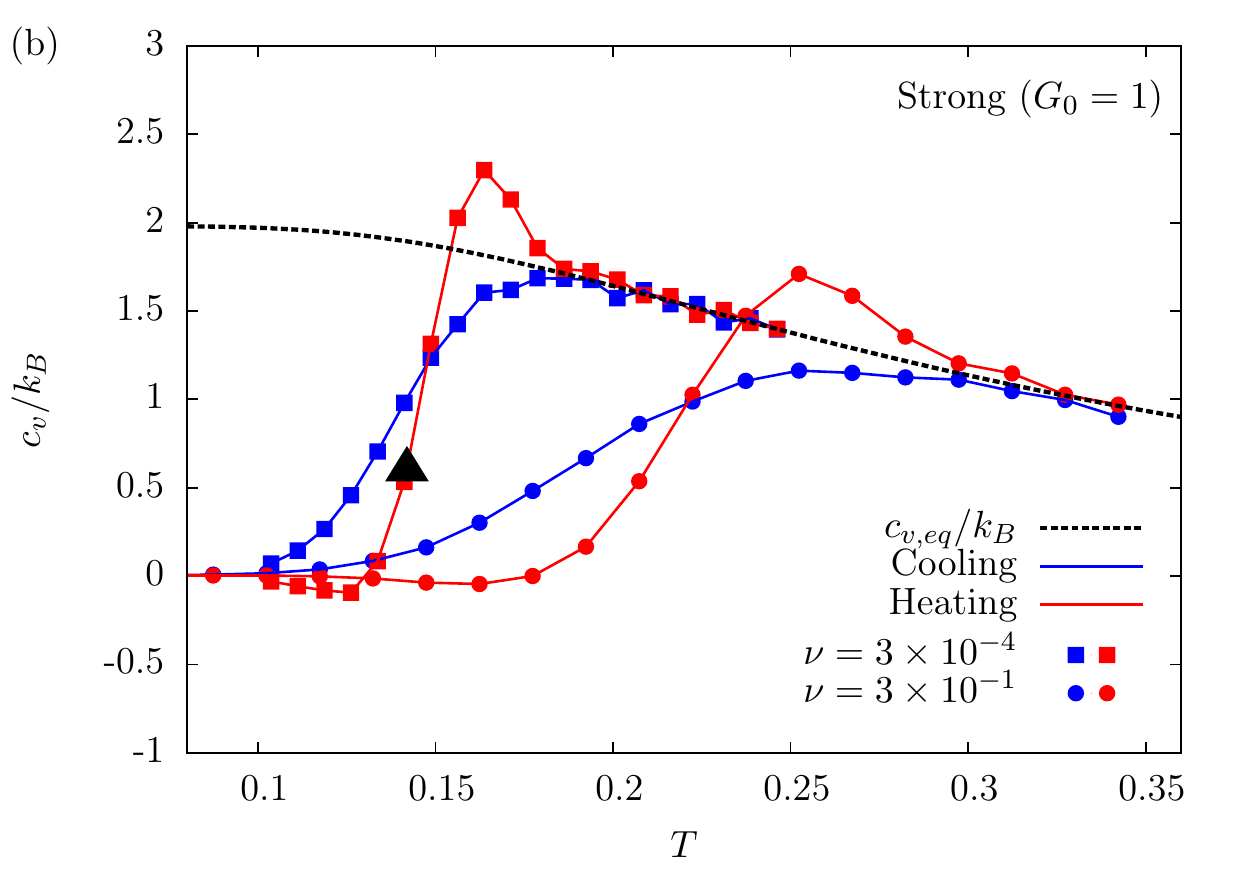}
	\caption{
		Specific heat capacity $c_v$ during a cooling-heating cycle for (a) fragile glass and (b) strong glass, with ${k_B=1}$.
		The black dashed line shows the equilibrium specific heat capacity $c_{v,eq} = d\varepsilon_{eq}/dT$, with $\varepsilon_{eq}$ is calculated from \eq{Eeq}.
		The black triangle marks the glass transition point measured from the heating data.
        Due to a lack of vibrations and the large $\nu$ used, $c_v$ and $c_{v,eq}$, which are more akin to heat capacity excess, are close to 0 at small $T$ and decrease rather significantly with $T$ at large $T$.
	}
	\label{fig:cvT}
\end{figure}

Figures~\ref{fig:eT} and \ref{fig:cvT} display typical kinetic Monte Carlo simulation results for $\varepsilon$ and $c_v$  for cooling/heating rates $\nu$ up to the slowest value $\nu = 3 \times 10^{-4}$ that we can simulate.
They successfully reproduce important features in experiments including energy and heat-capacity hysteresis with a prominent heat-capacity overshoot during heating \cite{li2017b,badrinarayanan2007,tropin2018book}. Nevertheless, we also observe that $c_v$ decreases more noticeably with $T$ at large $T$ than in experiments, which we attribute to a lack of particle vibrations and the large $\nu$ used in the simulations (see Sec. \ref{discussion:energy_and_heat_capacity_hysteresis}).
The definition of the glass transition temperature $T_g$ is given in Sec. \ref{section:Normalized_heat_capacity_and_Tg}.


\begin{figure}[t]
	\includegraphics[width=\linewidth]{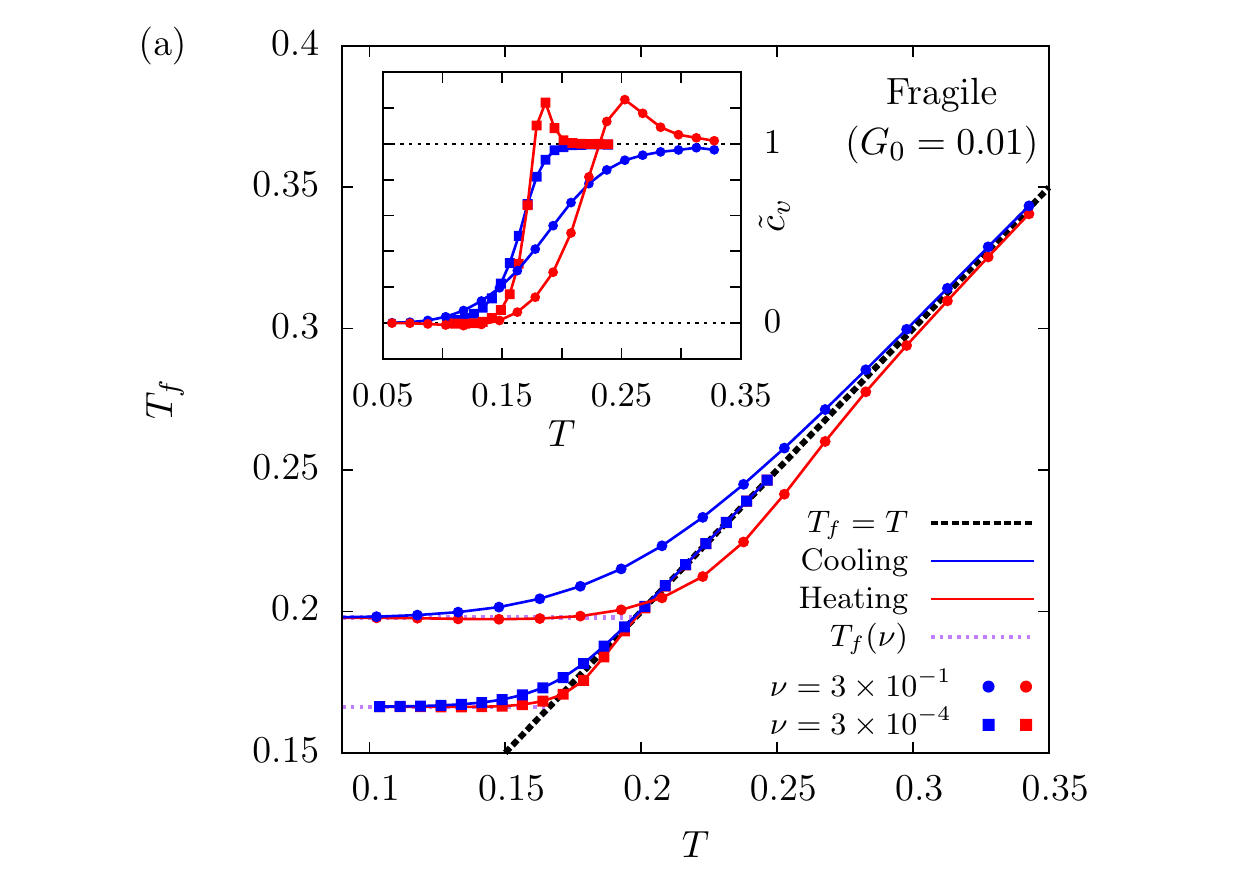}
	\includegraphics[width=\linewidth]{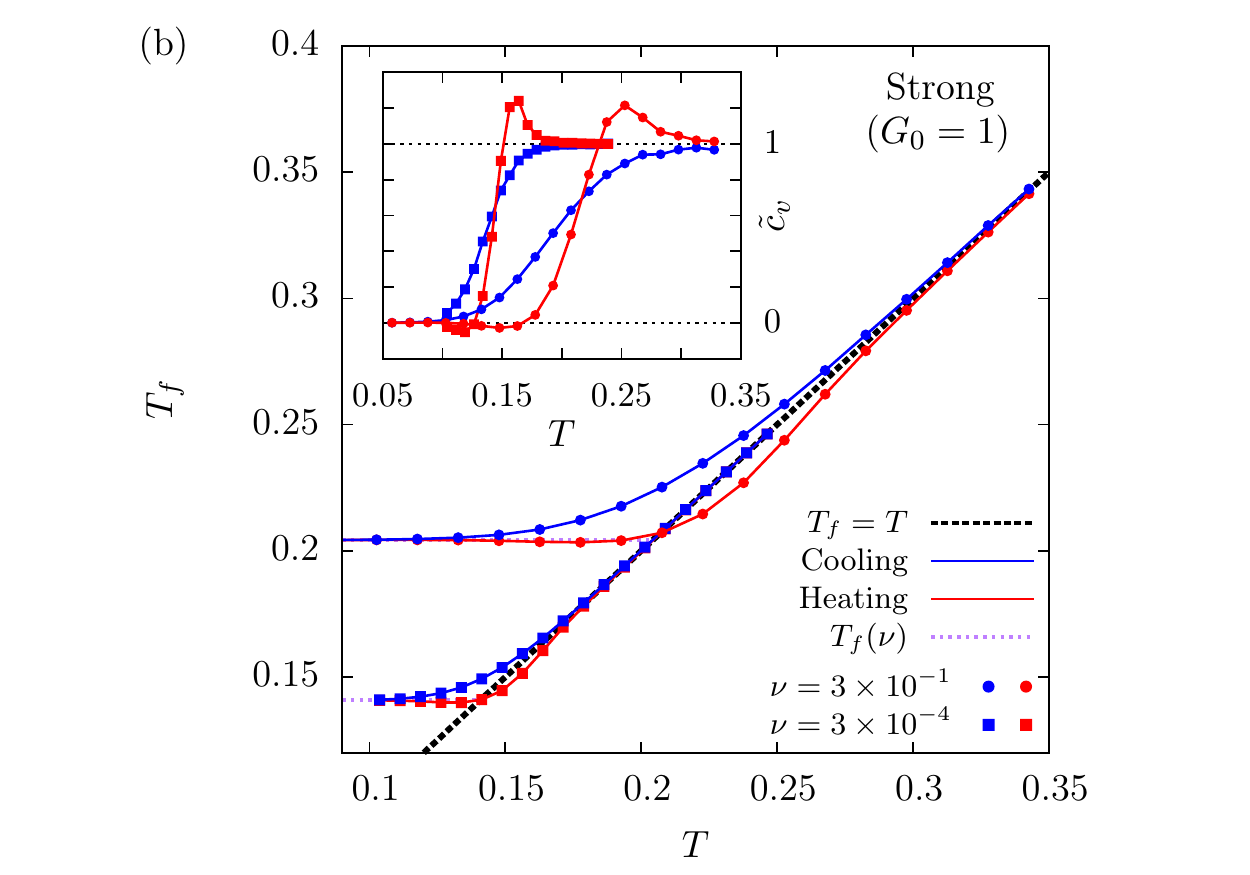}
	\caption{
		Plot of fictive temperature $T_{f}$ against bath temperature $T$ at cooling/heating rates $\nu$ for (a) fragile and (b) strong glasses.
		The black dashed line is the reference line $T_{f} = T$.
		The purple dotted lines indicate the fictive temperature $T_{f}(\nu)$ in the glass limit.
		Insets in both (a) and (b): Normalized heat capacity per particle $\tilde{c}_{v}$ versus bath temperature $T$ obtained using \eq{cv_norm}.
	}
	\label{fig:Tf}
\end{figure}

Most importantly, \fig{fig:cvT} shows large values of $c_v$ with a clear contrast between fragile and strong glasses. 
For $\nu=3 \times 10^{-4}$, $c_v$ shoots up in the heating process to nearly $12k_B$ for the fragile glass but only to about $2.5k_B$ for the strong glass. These peak values of $c_v$ occurring right above $T_g$ characterize the magnitudes of the heat-capacity jumps. 
We have expressed $c_v$ in unit of $k_B$, despite $k_B=1$, to highlight that the dimensionless quantity $c_v/k_B$ can be directly compared with experimental values. 
These values are of magnitudes similar to $c_v$ jumps of, for example, $8k_B$ and $1.6k_B$ for toluene \cite{ney2017} and a typical metallic glass \cite{ke2012}, 
which are fragile and moderately strong respectively.
The DPLM has thus provided $c_v$ jumps consistent with the experimental ones, which are significantly larger than those from conventional lattice models  \cite{fredrickson1986,mccullagh2005,nishikawa2020}.

\begin{figure}[t]
	\includegraphics[width=\linewidth]{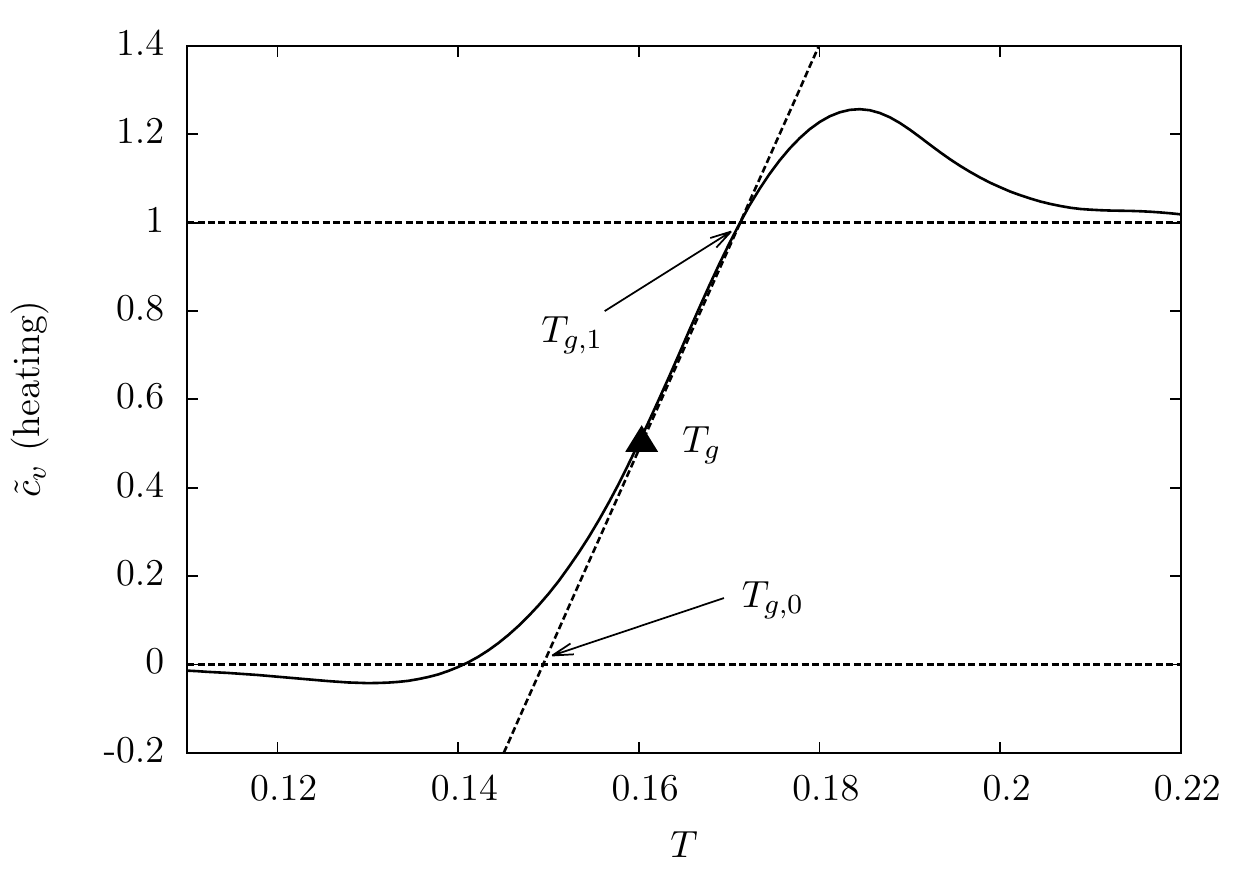}
	\caption{
		Schematic diagram showing the definition of $T_g$, $T_{g,0}$ and $T_{g,1}$.
	}
	\label{fig:Tg_from_cvnorm}
\end{figure}

\subsection{Fictive temperature and structural temperature}
We identify the fictive temperature $T_f$ of, in general, a non-equilibrium state with energy $\varepsilon$ as a numerically measurable structural temperature defined by \cite{lulli2020}
\begin{equation}
	\varepsilon = \varepsilon_{eq}(T_{f}),
	\label{Tf}
\end{equation}
where $\varepsilon_{eq}$ is the equilibrium energy, which is calculated analytically by using \eq{Eeq} (see the discussion section below) and is given in \fig{fig:eT} as black dashed line.
Further details on $\varepsilon_{eq}$ can be found in Appendix \ref{appendix:equilibrium_properties}.
Thus, $T_f$ measures the effective temperature of the particle interactions and reduces to the equilibrium temperature at equilibrium.
Note that its dependence on the particle configuration is explicitly known and is thus, strictly speaking, not a `fictive' quantity.  
\Fig{fig:Tf} plots $T_f$ against $T$ for different $\nu$ and $G_0$ during a cooling-heating cycle by using the same simulation results leading to \fig{fig:eT}.
We observe hysteresis in the evolution of $T_{f}$ analogous to that of $\varepsilon$ in \fig{fig:eT}. It also closely resembles hysteresis of $T_f$ observed in experiments \cite{tanaka2017}.

\subsection{Normalized heat capacity and $T_g$}
\label{section:Normalized_heat_capacity_and_Tg}
Besides particle energy and fictive temperature, the hysteresis can further be demonstrated by a normalized heat capacity per particle defined as 
\begin{equation}
\tilde{c}_{v} = \frac{c_{v}}{c_{v,eq}},
\label{cv_norm}
\end{equation}
where $c_{v,eq} = d\varepsilon_{eq}/dT$ is the equilibrium specific heat capacity.
Using the fictive temperature $T_f$ defined in \eq{Tf}, it can alternatively be expressed as 
\begin{equation}
\tilde{c}_{v} = \frac{dT_{f}}{dT},
\label{cv_norm2}
\end{equation}
a form more readily applicable to experiments  \cite{keys2013,li2017}.
The insets in \fig{fig:Tf} (a) and (b) show $\tilde{c}_{v}$ versus $T$ for the fragile  ($G_0 = 0.01$) and strong ($G_0 = 1$) glasses respectively. 
The results again closely resemble those observed in experiments \cite{keys2013,li2017}.
In particular, $\tilde{c}_{v}$ approaches 0 for  $T \ll T_g$ both in our simulations and in experiments.

In contrast to $c_v$, the hysteresis loops exhibited by $\tilde{c}_{v}$ for fragile and strong glasses closely resemble each other.
This suggests that the more pronounced hysteresis of $c_{v}$ for fragile glass mainly originates from the large value of $c_{v,eq}$ close to $T_g$.

We have adopted the glass transition temperature $T_g$ based on $\tilde{c}_{v}$ defined as the temperature at which $\tilde{c}_{v}=0.5$ as illustrated in \fig{fig:Tg_from_cvnorm}.
By drawing a tangent of $\tilde{c}_{v}$  at $T_g$, the onset temperature $T_{g,0}$ and the termination temperature $T_{g,1}$ of the glass transition can also be defined.

\begin{figure}[t]
	\includegraphics[width=\linewidth]{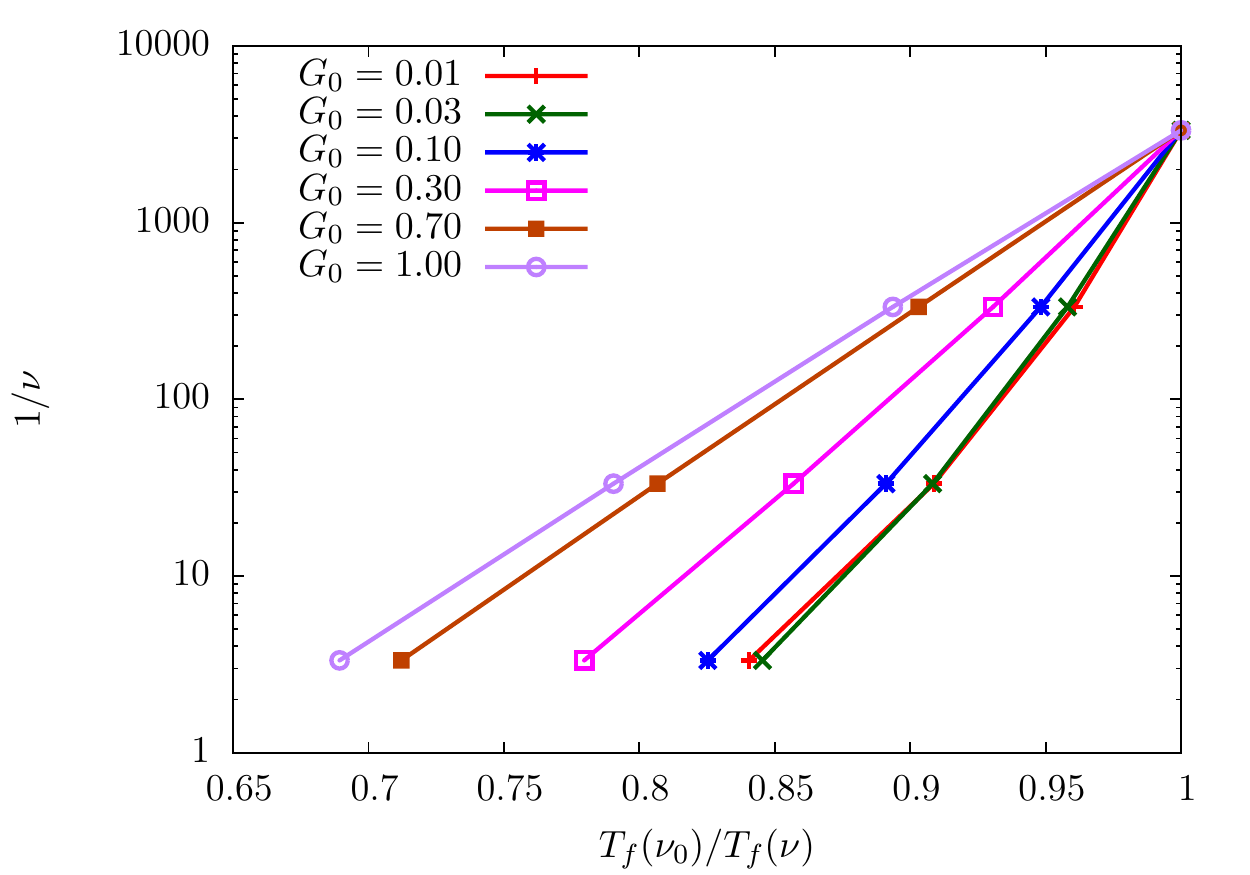}
	\caption{
		Angell plot of reciprocal cooling/heating rate $\nu^{-1}$ versus reciprocal fictive temperature  $T_{f}(\nu)^{-1}$ normalized by $T_{f}(\nu_0)$ where $\nu_0 = 3\times10^{-4}$.
	}
	\label{fig:Tfp_qc}
\end{figure}

\subsection{Angell plot based on cooling rate $\nu$}
Here, we measure at the end of cooling the fictive temperature $T_f(\nu)$, which is often considered close to $T_g$ at small $\nu$.
Based on the definitions as given from above, $1/\nu$ is a measure of the system relaxation time at temperature $T_f(\nu)$.
The results are thus displayed in the style of an Angell plot in \fig{fig:Tfp_qc}, where $1/\nu$ is plotted against $T_{f}(\nu_0)/T_f(\nu)$ with $\nu_0= 3\times 10^{-4}$.
%
Results are similar to previous studies with both Arrhenius \cite{moynihan1974} and super-Arrhenius \cite{yue2004} behaviors have been observed.

\begin{figure}[t]
	\includegraphics[width=\linewidth]{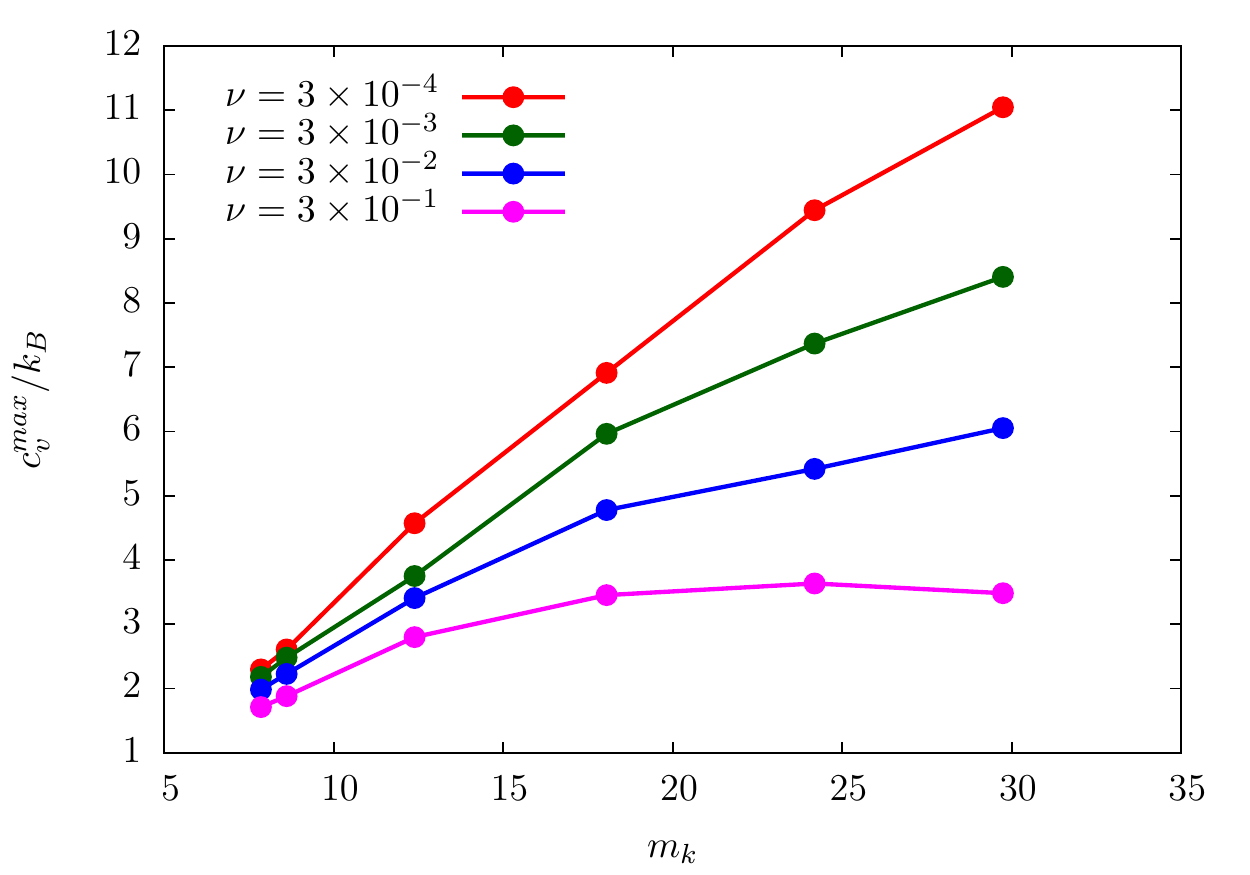}
	\caption{
		Correlation between peak value $c^{max}_v$ of $c_v$ during heating and kinetic fragility $m_k$. 
	}
	\label{fig:cvpeakmk}
\end{figure}

\subsection{Heat-capacity overshooting magnitude}
The $c_v$ hysteresis as shown in \fig{fig:cvT} is more pronounced for the fragile than for the strong glass in agreement with experiments \cite{tropin2018book,li2017b}.
The correlation is further quantified in Fig.~\ref{fig:cvpeakmk}, where the maximum value of $c_v$ during overshoot in the heating process, denoted by $c^{max}_{v}$, is plotted versus the kinetic fragility index $m_k$ at various heating rates $\nu$.
Note that $m_k = \partial\log\tau/\partial(T^{*}_g/T)\vert_{T^{*}_{g}}$ is calculated using data from \fig{fig:Angelltau}.
We have used a reference relaxation time of $\tau_r=10$ to define the glass transition temperatures $T^{*}_{g}$, which is about the longest time scale we can simulate but is indeed small when compared to experiments (see Appendix \ref{appendix:glassy_dynamics} for further details).
This leads to $m_k$ much smaller than the experimental ones, as explained in detailed in \Ref{lee2020}.
From \fig{fig:cvpeakmk}, 
$c^{max}_v$ is seen increasing with $m_k$.
It is also observed that the heating rate affects $c^{max}_v$ for fragile glass more than from strong glass, a feature that has been observed in experiments \cite{tropin2018book}.

\begin{figure}[t]
	\includegraphics[width=\linewidth]{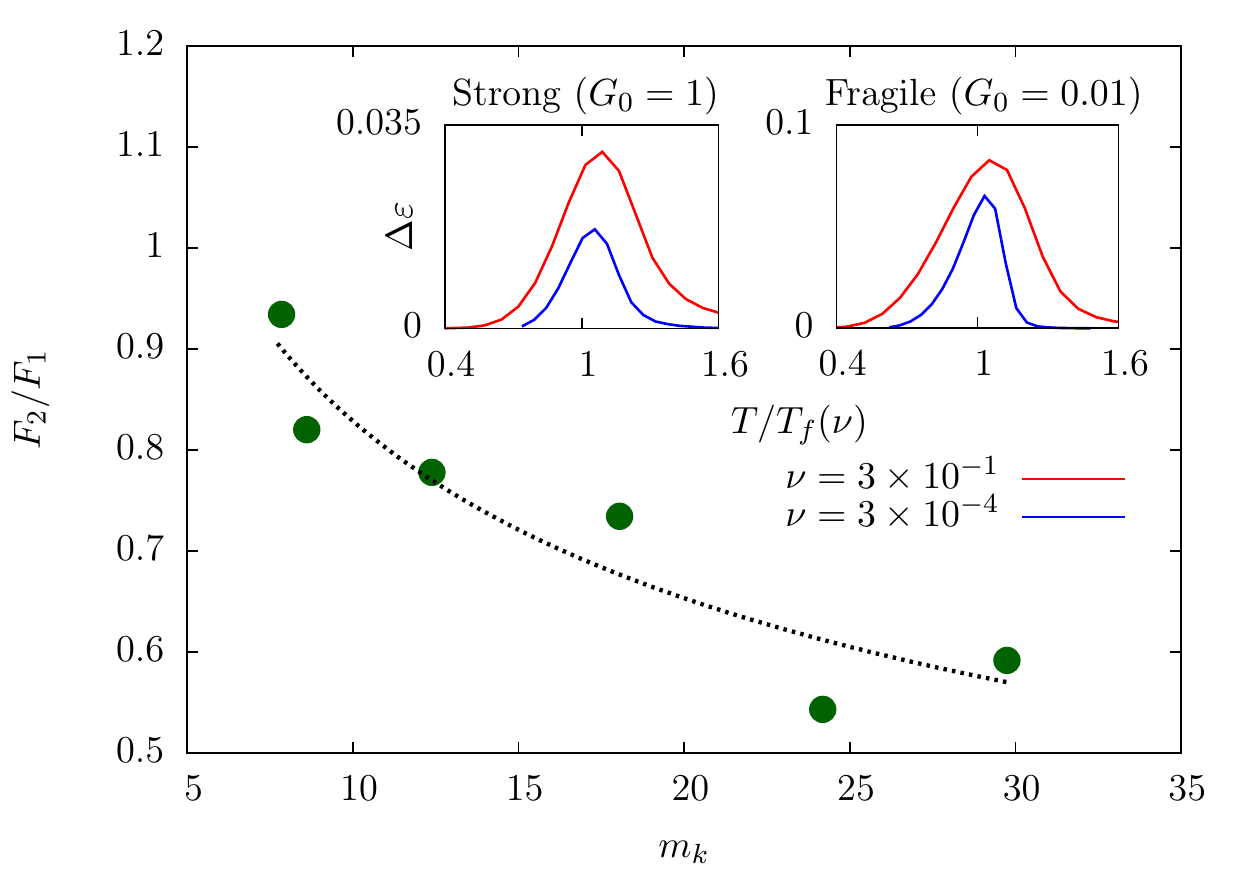}
	\caption{
		Asymmetric factor $F_2/F_1$ against kinetic fragility $m_k$, at cooling rate $\nu = 3\times 10^{-4}$.
		Insets: Energy difference per particle $\Delta \varepsilon$ against rescaled temperature $T/T_{f}(\nu)$ for (left panel) strong glass ($G_0 = 1$) and (right panel) fragile glass ($G_0 = 0.01$).
	}
	\label{fig:relating_mk}
\end{figure}

\subsection{Asymmetry in Hysteresis loop}

We now study the energy difference per particle $\Delta \varepsilon$ at temperature $T$ during cooling compared with heating following \Ref{li2017}.
The insets of \fig{fig:relating_mk} plot $\Delta \varepsilon$ against $T/T_f(\nu)$ for strong ($G_0 = 1$) and fragile ($G_0 = 0.01$) glasses, at $\nu = 3\times 10^{-1}$ and $3\times 10^{-4}$, where $\Delta \varepsilon$ is obtained by subtracting $\varepsilon$ from cooling by that from heating in \fig{fig:eT}.
In each case, we observe a peak which is in general skewed.  
The skewness can be quantified by an asymmetric factor $F_2/F_1$, where $F_1$ is the left half width at half maximum (HWHM) of the peak while $F_2$ is the right HWHM.
The main figure of \fig{fig:relating_mk} plots $F_2/F_1$ against the kinetic fragility $m_k$ for $\nu = 3\times 10^{-4}$, which qualitatively resembles experimental results \cite{li2017}. 
The asymmetry arises from the highly nonlinear temperature dependence of the relaxation dynamics, which can be analyzed using the Tool–Narayanaswamy–Moynihan–Hodge (TMNH) equations as shown in \Ref{li2017}.
Our results show that the DPLM is able to naturally reproduce the trend of a stronger hysteresis asymmetry of the fragile glasses compared with strong glasses.

\section{Discussions}

\subsection{Energy and heat capacity hysteresis}
\label{discussion:energy_and_heat_capacity_hysteresis}
Using the DPLM, we have reproduced the energy and heat capacity hysteresis during cooling-heating cycles typical of glasses as shown in \figs{fig:eT}{fig:cvT}.  While the main features of the these hysteresis loops are captured, there are also some discrepancies, which we attribute to a lack of particle vibrations and the large $\nu$ used in the simulations.

First, $c_v$ from \fig{fig:cvT} is much closer to 0 at small $T$ than in experiments. This is easily understandable as the DPLM does not simulate particle vibrations. A particle configuration corresponds to an inherent structure and $\varepsilon$ represents the configurational energy \cite{deng2019}.
In the glass phase at $T \ll T_g$ with frozen configurations, the particle energy $\varepsilon$ thus approaches a constant resulting at $c_v\simeq 0$. In fact, $c_v$ in the DPLM can better be compared with heat capacity excess from experiments, which is also close to zero in the glass phase \cite{angell2008review}. 

Second, we observe that $c_v$ decreases more noticeably with $T$ at large $T$ than in experiments. This results from a similar  property of the equilibrium heat capacity $c_{v,eq}$. Due to the lack of vibrations in the DPLM, the particle energy $\varepsilon$ attains a finite limit as $T\rightarrow\infty$, similar to the case of typical lattice models in statistical physics. This implies a diminishing $c_v$ at large $T$, in contrast to typically molecular systems. An additional factor is that the adopted heating/cooling rates $\nu$ are many orders larger than the experimental range. For example, for the fragile glass at $\nu = 3 \times 10^{-4}$, hysteresis occurs over $T$ ranging from $T\simeq 0.13$ to 0.20, leading to a  width $\Delta T \simeq 0.07$ of the hysteresis as observable in \fig{fig:cvT}(a). This  width is about 43\% of $T_g=0.163$ and this ratio decreases as $\nu$ decreases. In contrast, the width of the hysteresis loops extends over only about 10\% of $T_g$ in experiments due to the much lower cooling/heating rates \cite{tropin2018book}. Because of the much wider temperature range covered in our simulations, we observe from \fig{fig:cvT}(a) a noticeable continuous decrease of $c_v$ with $T$ beyond the hysteresis, whilst $c_v$ appears to approach a constant in experiments. We thus expect these different features between simulations and experiments to diminish if a much slower $\nu$ can be used, which however is impractical computationally. 

The hysteresis phenomenon observed here is similar to  those in 
typical systems with finite response times and can be modeled for example by 
the Tool-Narayanaswamy-Moynihan theory \cite{hodge1994}.
The process can be understood as follows.
At the beginning of the cooling process when $T$ is high, the system equilibrates fast with a short structural relaxation time $\tau$, and the energy per particle $\varepsilon$ closely follows the equilibrium value $\varepsilon_{eq}$.
As $T$ decreases, $\tau$ increases.
Following Deborah's condition \cite{hodge1994}, when $T$ becomes so low that $|d \tau / d t|=1$, i.e. $|d \tau / dT| = \nu^{-1}$, 
the system cannot fully equilibrate and falls out of equilibrium. For slower (faster) cooling, this takes place at lower (higher) $T$. 
In the non-equilibrium state, the system partially retains its preceding state, which is the higher-temperature near-equilibrium state, leading to $\varepsilon > \varepsilon_{eq}$.
The discrepancy $\varepsilon - \varepsilon_{eq}$ widens as $T$ decreases.
When $T$ drops to such a low temperature that $| d \tau / dt | \gg 1 $, structural relaxation can hardly happen and $\varepsilon$ freezes. 

In contrast, at the beginning of the heating process, the system has a longer $\tau$ inherent from its non-equilibrium state at lower temperature, leading to $\varepsilon$ less than the previous value at the same $T$ during cooling.
This originates the observed hysteresis, which closes only at a temperature high enough so that $| d \tau / dt | \ll 1$. 

\subsection{Heat capacity jump}
As aforementioned, the correlation reproduced above between $c_v$ jump and fragility is mainly caused by equilibrium properties of the glasses. 
Note that from \fig{fig:cvT}, $c_{v,eq}\simeq c_v> 0$ well above $T_g$ and $c_{v,eq}\simeq c_v \simeq 0$ well below $T_g$. The magnitude of $c_{v,eq}$ close to $T_g$ basically dictates the jump of $c_v$. The 
contrast of $c_v$ between fragile and strong glasses therefore reduces to a similar contrast in $c_{v,eq}$.
Equilibrium thermodynamics stipulates that
\begin{equation}
  c_{v,eq} = T\frac{d S}{d T}
  \label{cveq}
\end{equation}
under constant volume conditions, where $S$ denotes the entropy per particle.   
Before a quantitative analysis, it is immediately understandable from \eq{cveq} why a fragile glass has a large $c_{v,eq}$, and thus a large $c_v$ jump.
Specifically, as $T$ decreases towards $T_g$, the entropy $S$ of fragile glasses have been shown to admit a dramatic drop, which is associated with increasingly constrained kinetic pathways characteristic of the glass transition \cite{lee2020}. This precisely implies a large $dS/dT$ close to $T_g$ and thus, using \eq{cveq}, also a large $c_{v,eq}$ and $c_v$.


Furthermore, it is instructive to compare the magnitude of $c_{v,eq}$ with naive predictions from  
equipartition of energy, which is exact for harmonic inter-molecular potentials. In the DPLM,
realized interaction $V_{s_i s_j}$ between neighboring sites $i$ and $j$ in \eq{E} is time dependent because $s_i$ and $s_j$ change as particles move around.
Each $V_{s_i s_j}$ is hence a degree of freedom of the system. If its distribution takes a simple unimodal form close to that in a harmonic oscillator, equipartition of energy suggests an average interaction of ${\sim}k_B T/2$ above $V_0$, leading to a heat capacity of $k_B/2$ per interaction.  Assuming a small void density $\phi_v$, we get $c_{v,eq} \simeq z k_B/4= k_B$, where $z = 4$ is the lattice coordination number.
A $c_{v,eq}$ much larger than $k_B$ for a fragile glass therefore requires that the distribution of $V_{s_i s_j}$ must deviate drastically from a unimodal form, as in a bi-component distribution, and this will be further explained below.
Note that this estimate is in general distinct from $(d/2)k_B$ from the Dulong-Petit law, where $d$ is the spatial dimension.

\begin{figure}[t]
	\includegraphics[width=\linewidth]{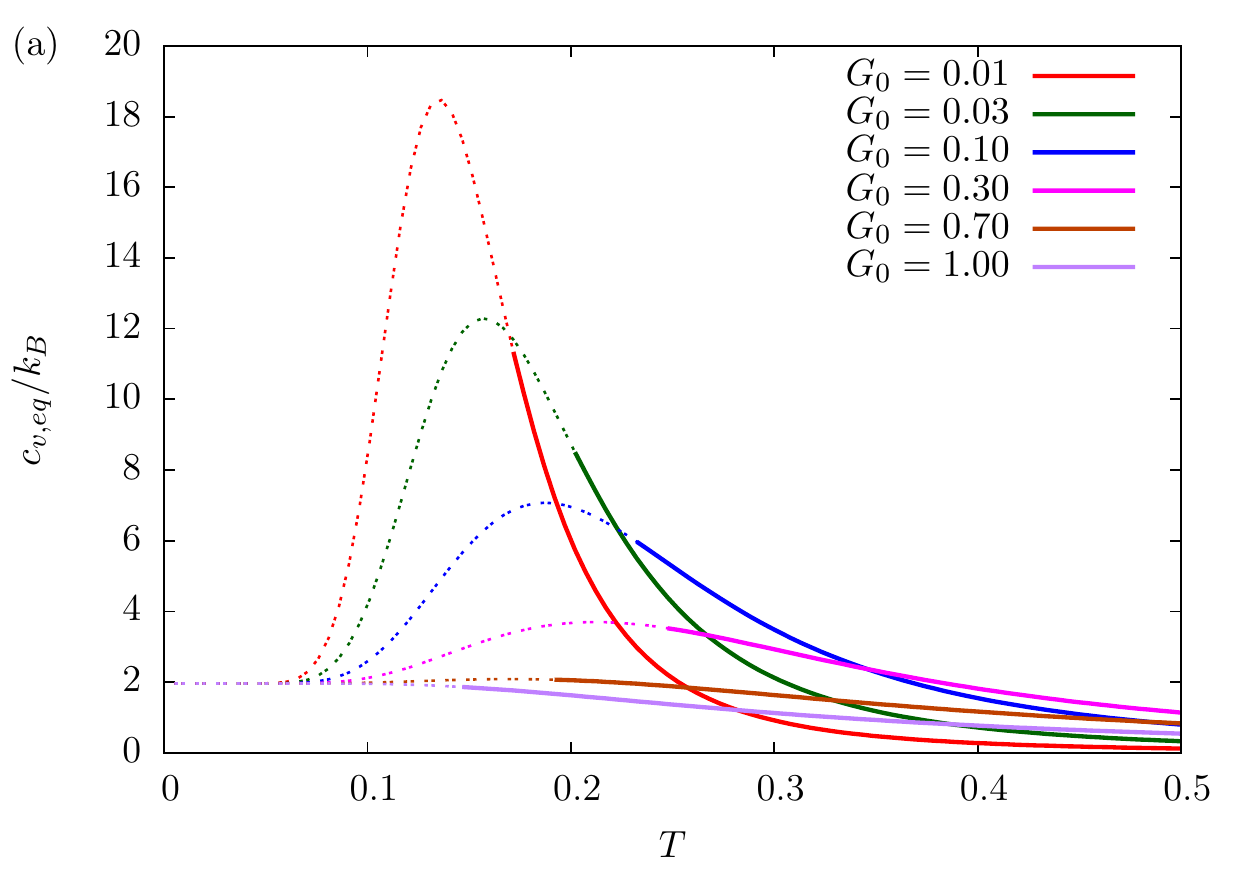}
	\includegraphics[width=\linewidth]{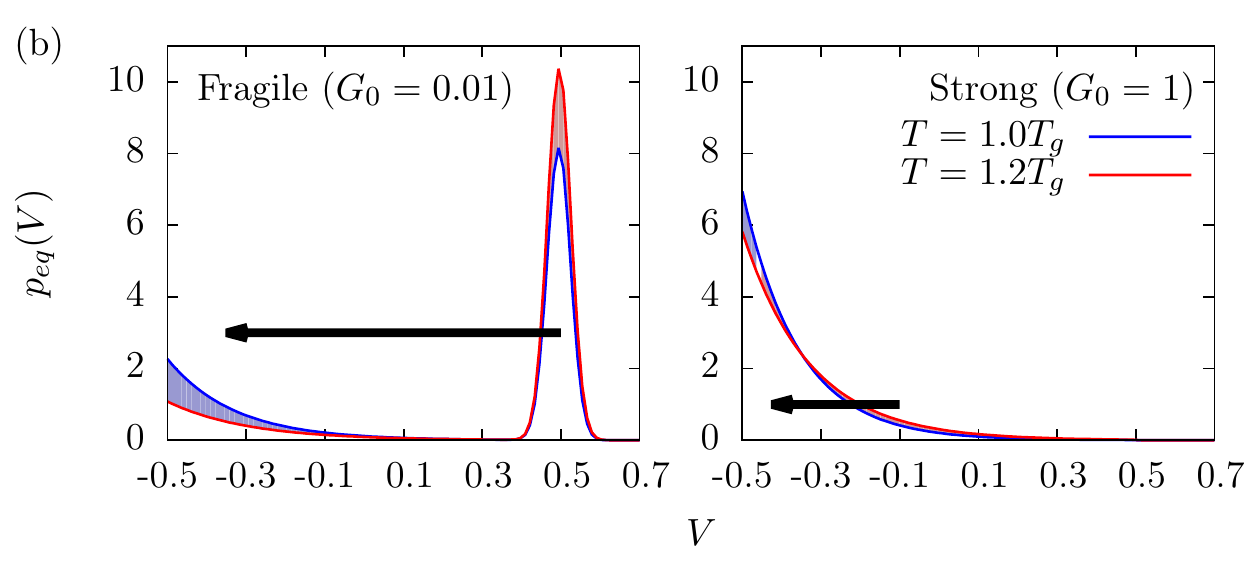}
	\caption{
		(a). The temperature dependence of equilibrium specific heat capacity $c_{v,eq}(T)$ for various $G_0$.
		Physically unobservable data for $T < T_g$ are plotted with dotted lines.
		(b). Equilibrium distribution $p_{eq}(V)$ of interaction energy $V$ at $T_{g}$ (blue line) compared with  that at $1.2T_{g}$ (red line) for fragile glass (left panel) and strong glass (right panel).
		The loss in spectral weight of the high-energy interactions (red area) is balanced by the gain of the low-energy ones (blue area). Such weight transfer (black arrow) is much more substantial in fragile than in strong glass.
		For the ease of illustration, we have smoothed $g(V)$ in \eq{gV} by replacing the uniform and the delta components by simple Fermi and Gaussian functions respectively. 
	}
	\label{fig:cveqT}
\end{figure}

Equilibrium properties of the DPLM including $c_{v,eq}$ will now be analytically calculated.
As derived in \Ref{zhang2017} and extensively verified numerically \cite{zhang2017,lulli2020,lee2020}, particles in the DPLM arrange themselves at equilibrium in such a way that the realized interaction energy $V_{s_i s_j}$ follows exactly the $a$ $posterior$ distribution
$	p_{eq}(V) = \frac{1}{\mathcal{N}} g(V) \exp(-V/k_BT)$,
where $\mathcal{N} = \int dV g(V) e^{-V/k_BT}$ is a normalization factor.
The equilibrium energy per particle is thus
\begin{equation}
	\varepsilon_{eq} = \frac{z}{2}\int dV ~ V p_{eq}(V)
	\label{Eeq}
\end{equation}
at small void density $\phi_v$.
One then finds the equilibrium specific heat capacity by
$c_{v,eq} = d \varepsilon_{eq}/dT$, which is a more convenient expression than \eq{cveq}.
With $g(V)$ given by Eq.~(\ref{gV}), $\varepsilon_{eq}$ and $c_{v,eq}$ can be explicitly worked out (see Appendix \ref{appendix:equilibrium_properties}), as already 
plotted in Figs.~\ref{fig:eT} and \ref{fig:cvT}.

\Fig{fig:cveqT}(a) compares $c_{v,eq}$ for various values of $G_0$. We observe that $c_{v,eq}$ at $T_g$ decreases monotonically with $G_0$. At the small $T$ limit, $c_{v,eq}$ converges to $zk_B/2$ independent of $G_0$. This is because the low-energy uniform component of $g(V)$ dominates, leading to effectively a unimodal situation with 
$\varepsilon_{eq} \approx (z/2) (k_BT+V_0)$.
This leads to $c_{v,eq} \simeq zk_B/2$ which differs from the equipartition prediction $zk_B$ explained above only by a factor of 2. 
For the strong glass,  $z k_B/2$ directly approximates the $c_v$ jump.

\subsection{Two-state picture}
The increasingly prominent peak of  $c_{v,eq}$ in {\fig{fig:cveqT}(a) as $G_0$ decrease may seem to suggest an underlining criticality. However, there is no  divergence at any finite $G_0$. Instead, the peak characterizes $T$ at which the relative importance of the two components of $g(V)$ in the two-state picture depends most sensitively on $T$.}
According to \eq{Eeq}, the $T$ dependence of the particle energy $\varepsilon_{eq}$ can be ultimately traced to a spectral weight transfer in the $a$ $posterior$ distribution $p_{eq}(V)$ from high-energy interactions to low-energy ones.
This is illustrated in \fig{fig:cveqT}(b), where we compare $p_{eq}(V)$ at $1.2T_g$ with that at $T_g$. For the strong glass in \fig{fig:cveqT}(b) (right panel), a small probability weight is transferred to interactions with energies lowered on average by about $0.35$ (black arrow). In sharp contrast, for the fragile glass in \fig{fig:cveqT}(b) (left panel), the transfer is over an energy difference of about 0.9 (black arrow) and the weight of the low-energy part nearly doubles.
Note that this contrast does not result from distinct energy scales, characterizable for example by $k_B T_g$ which indeed take similar values of $0.149$ and $0.163$ respectively for the strong and fragile glasses.
The significant transfer for the fragile glass occurs due to a competition between entropy that favors the high-energy component of $g(V)$ and the Boltzmann factor that favors its low-energy part, noting that the high-energy component has a much higher entropy due to its large weight of $1-G_0=0.99$ compared with the weight $G_0=0.01$ of the low-energy part.
Such a drastic spectral transfer is only possible due to the bi-component form of $g(V)$ highly relevant to fragile glasses, and is absent for the essentially unimodal form for strong glasses.
Note that the transfer also causes the kinetic slowdown in fragile glasses \cite{lee2020} so that $T_g$ occurs where $c_{v,eq}$ varies sharply.

\section{Conclusion}
To conclude, using the DPLM, we have reproduced the major experimentally observed features of the heat capacity hysteresis of glass formers: the large value of $c_v$ and the strong correlation with fragility. The large $c_v$ jump of fragile glass during cooling below the glass transition temperature is demonstrated to inherent from the large equilibrium value of $c_v$.
Based on a two-state picture, the latter is shown to be controlled in turn by a crossover from a high-energy interaction state to a low-energy one, a process which also induces the high fragility.
Our work shows that particle models defined on a lattice, in contrast to defect models, are capable of capturing glass thermodynamics intrinsically, with the essential physics intuitively revealed.  

Note that using only the two-state model, one can already study simple kinetics by postulating superposition rules \cite{rui2018}, but not complex dynamical phenomena. 
On the other hand, the DPLM can reproduce a wide range of characteristic glassy dynamics \cite{zhang2017,lee2020} and has also been successfully employed to address 
Kovacs paradox \cite{lulli2020} and Kovacs effect \cite{lulli2019} on glass aging.
Our approach successfully combines the DPLM with the two-state picture, so that both thermodynamics and kinetics of strong and fragile glasses can be studied microscopically under a consistent set of assumptions.
Investigating the rich phenomena exhibited by a diverse range of various glasses in such a unified manner should be of particular importance.


\begin{acknowledgments}
We thank  helpful discussions with R. Shi and P. Sollich.
This work was supported by General Research Fund of Hong Kong (Grant 15303220) and
National Natural Science Foundation of China (Grant 11974297).
\end{acknowledgments}

\appendix

\section{Equilibrium properties}
\label{appendix:equilibrium_properties}

\subsection{Equilibrium statistics}
A surprising feature of the DPLM is that it has exactly solvable equilibrium statistics \cite{zhang2017}, predictions of which have been 
extensively and accurately verified numerically \cite{zhang2017,lulli2020,lee2020}.
For the canonical ensemble considered in our DPLM simulations, the partition function $Z$ in the thermodynamic limit is given by \cite{zhang2017,lee2020}

\begin{equation}
	\frac{Z}{N!} = \sum_{\bracebk{n_i}} e^{-\beta N_{b}U},
	\label{Z2}
\end{equation}
where $n_i=0$ or 1 is the occupancy at site $i$ and
$\beta = 1/k_{B}T$.
Also, $N_b$ denotes the number of nearest neighboring particle pairs and $U$ is the free energy of a pair interaction defined by
\begin{equation}
	U = -\frac{1}{\beta}\ln{\mathcal{N}},
	\label{U}
\end{equation}
where
\begin{equation}
	\mathcal{N} = \int dV g(V) e^{-\beta V}.
	\label{N}
\end{equation}
Interaction $V_{s_{i}s_{j}}$ realized in the system at any instance follows the  $a$ $posterior$ distribution 
\begin{equation}
	p_{eq}(V) = \frac{1}{\mathcal{N}} g(V) \exp(-\beta V).
	\label{peq}
\end{equation}

Based on the above equilibrium statistics, we now calculate the average energy and heat capacity. 
The equilibrium energy per particle $\varepsilon_{eq}$ at small void density $\phiv$ can be calculated from \eq{Eeq}, which can be rewritten as
\begin{equation}
	\varepsilon_{eq} = \frac{z}{2}\frac{\partial \ln{\mathcal{N}^{-1}}}{\partial \beta}.
	\label{Eeq2}
\end{equation}
The equilibrium heat capacity $c_{v,eq}$ is then obtained by taking the temperature derivative of \eq{Eeq2}, which 
is given by
\begin{equation}
	c_{v,eq} = \frac{z}{2}k_B\beta^{2}\frac{\partial^{2}\ln{\mathcal{N}}}{\partial\beta^{2}}.
	\label{cveq2}
\end{equation}
Thus, the problem is reduced to calculating $\mathcal{N}$.
For the bi-component form of $g(V)$ in \eq{gV}, \eq{N} gives
\begin{equation}
\mathcal{N} = \frac{G_0}{\beta\Delta V}(e^{-\beta V_0} - e^{-\beta V_1}) + (1 - G_0)e^{-\beta V_1}.
\label{N_uni_delta}
\end{equation}
Using \eqr{Eeq2}{N_uni_delta}, $\varepsilon_{eq}$ and $c_{v,eq}$ can be evaluated numerically.

For various values of $G_0$, $c_{v,eq}$ is plotted in \fig{fig:cveqT}.
We observe a strong dependence of $c_{v,eq} $ on $G_0$ and thus on the fragility.
In particular, for the most fragile glass studied here with $G_0 = 0.01$, $c_{v,eq}$ shoots up to about $18 k_B$, where $k_B= 1$.
In sharp contrast, for the strong glass with $G_0 = 1$, $c_{v,eq} \leq 2 k_B$.


\begin{figure}[t]
	\includegraphics[width=\linewidth]{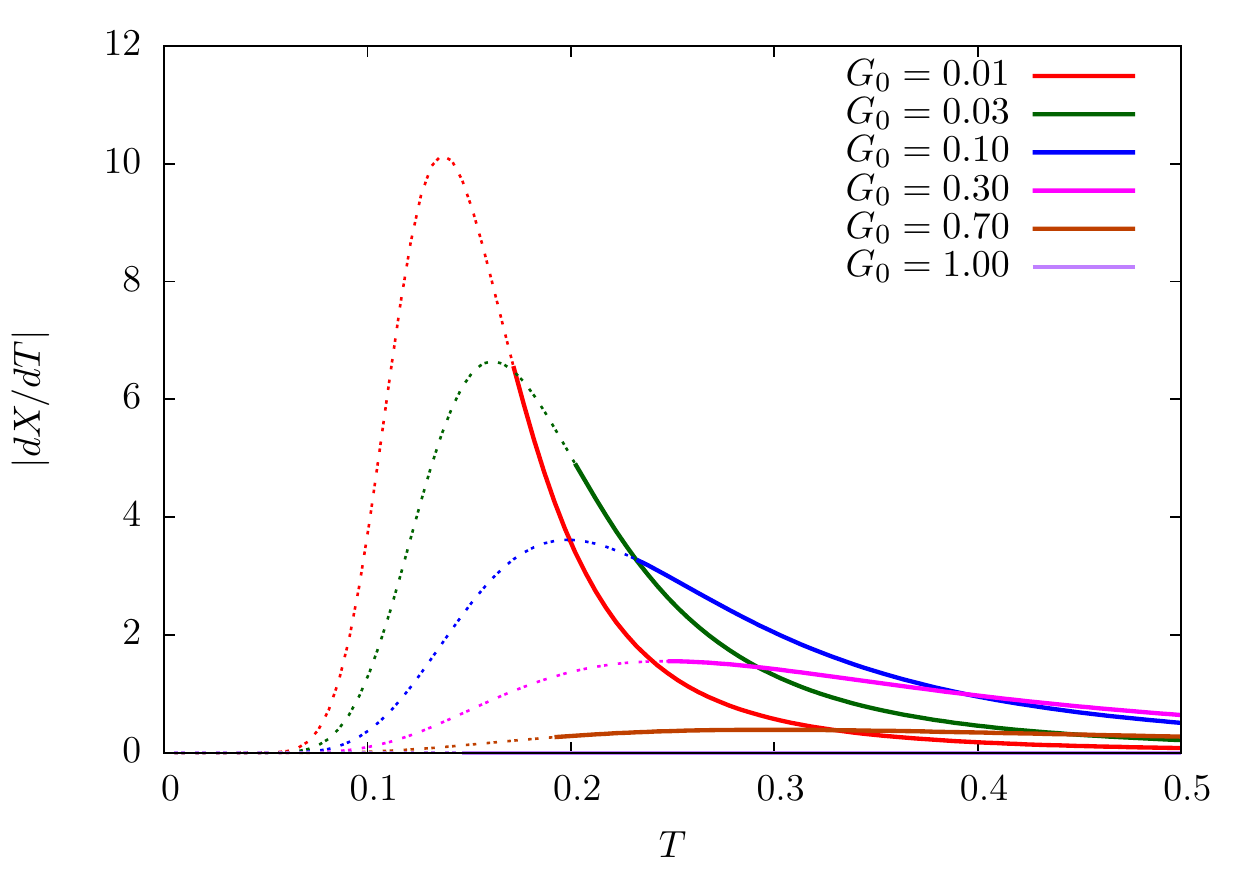}
	\caption{
		Plot of $| dX/dT | = -dX/dT$ against temperature $T$ for various $G_0$, where $X$ is the probability weight of the low-energy component of the particle interaction distribution.
		Regions for $T < T_g$ are plotted with dotted lines, where $T_g$ is the glass transition defined in \fig{fig:Tg_from_cvnorm}.
	}
	\label{fig:dXdT}
\end{figure}

\subsection{Bi-component $g(V)$ and its two-state nature}
For a better intuitive understanding of the large peak value of $c_{v,eq}$ for fragile glass, we first derive further analytic expressions of $c_{v,eq}$. 
Following \Ref{lee2020}, the bi-component interaction distribution $g(V)$ in \eq{gV}  can be rewritten as
\begin{equation}
g(V) = g_{A}(V) + g_{B}(V),
\label{gV2}
\end{equation}
with the labels A and B specifying the two components, i.e.
\begin{eqnarray}
g_{A}(V) &=& \frac{G_0}{\Delta V}, \label{gV_A}\\
g_{B}(V) &=& (1 - G_0)\delta(V - V_1), \label{gV_B}
\end{eqnarray}
where $V \in [V_0, V_1]$ with $V_0 = -0.5$, $V_1 = 0.5$ and $\Delta V = V_1 - V_0 = 1$.

The probabilistic weight of the uniform component, i.e. component A, equals
\begin{equation}
X = \frac{\mathcal{N}_{A}}{\mathcal{N}_{A} + \mathcal{N}_{B}}, \label{X} \\
\end{equation}
while the weight for the Dirac distribution component is $1 - X$, with  
\begin{eqnarray}
\mathcal{N}_{A,B} &=& \int dV g_{A,B}(V) e^{-V/k_BT}. \label{N_AB} 
\end{eqnarray}
Limiting to any one of the components, the equilibrium interaction distribution is
\begin{eqnarray}
p^{A,B}_{eq} &=& \frac{1}{\mathcal{N}_{A,B}} g_{A,B}(V)e^{-V/k_BT}. \label{peq_AB}
\end{eqnarray}
The average interaction energy within each component is simply computed by
\begin{equation}
\overline{V}_{A,B} = \int dV ~V p^{A,B}_{eq}(V).
\label{V_AB}
\end{equation}
Inserting \eqs{gV_A}{gV_B} into \eqs{N_AB}{V_AB}, we have \cite{lee2020}
\begin{eqnarray}
\mathcal{N}_{A} &=& \frac{G_0k_BT}{\Delta V}(e^{-V_0/k_BT} - e^{-V_1/k_BT}), \label{N_A} \\
\mathcal{N}_{B} &=& (1 - G_0)e^{-V_1/k_BT}, \label{N_B} \\
\overline{V}_{A} &=& V_0 + k_BT - \frac{\Delta V}{e^{\Delta V/k_BT} - 1}, \label{V_A} \\
\overline{V}_{B} &=& V_0 + \Delta V. \label{V_B}
\end{eqnarray}

\begin{figure}[t]
	\includegraphics[width=\linewidth]{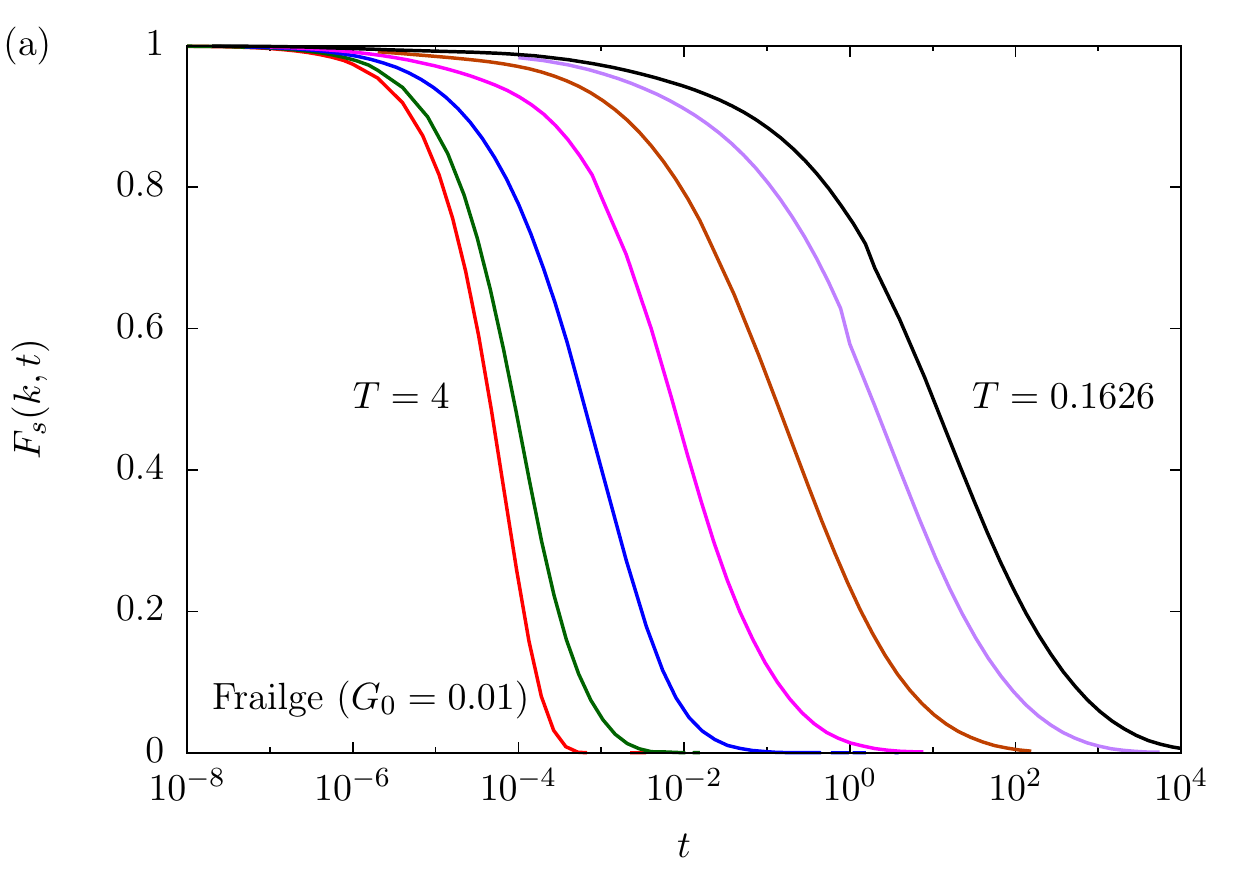}
	\includegraphics[width=\linewidth]{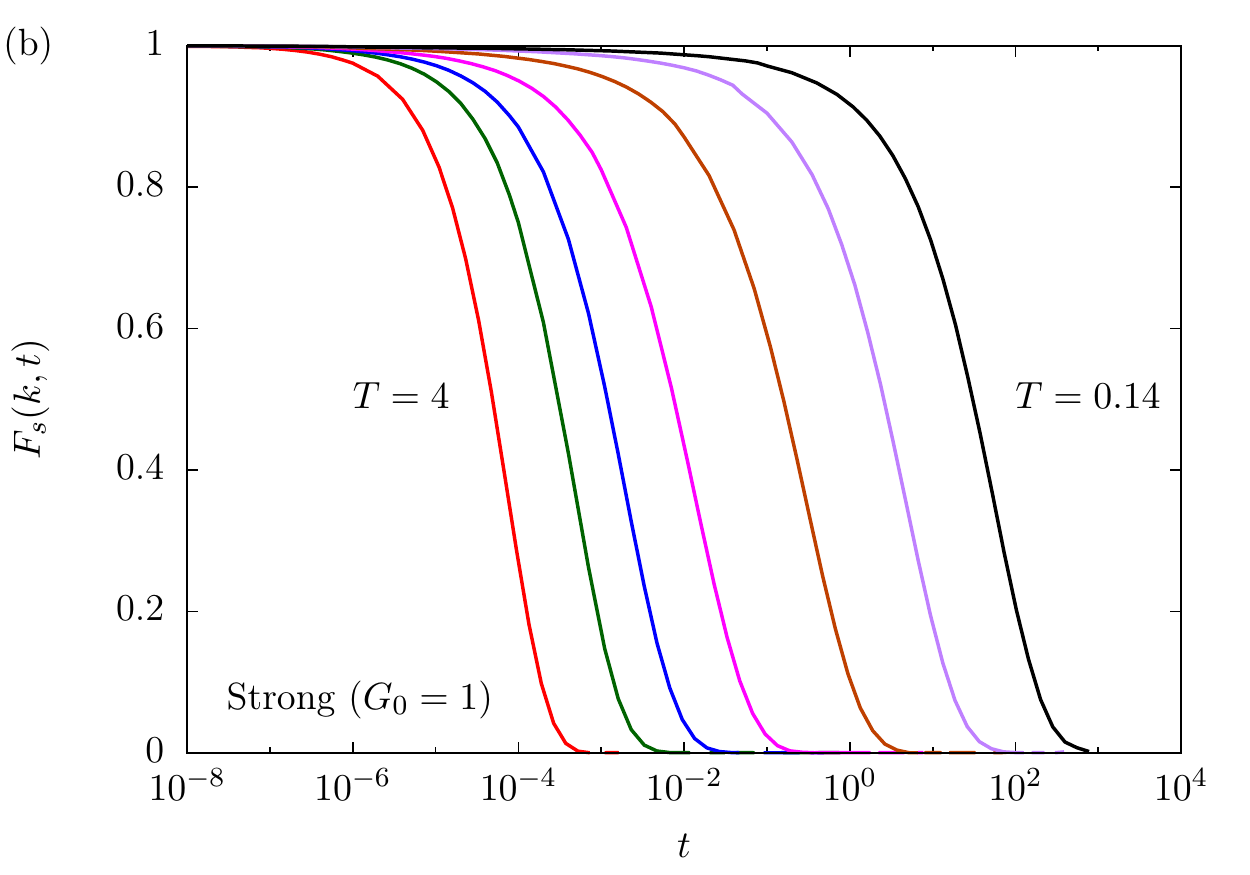}
	\caption{
		Self-intermediate scattering function $F_{s}(k, t)$ against time $t$ for (a) fragile and (b) strong glasses with $k = 2\pi/\lambda$ and $\lambda = 2$.
	}
	\label{fig:SIF}
\end{figure}
\begin{figure}[h]
	\includegraphics[width=\linewidth]{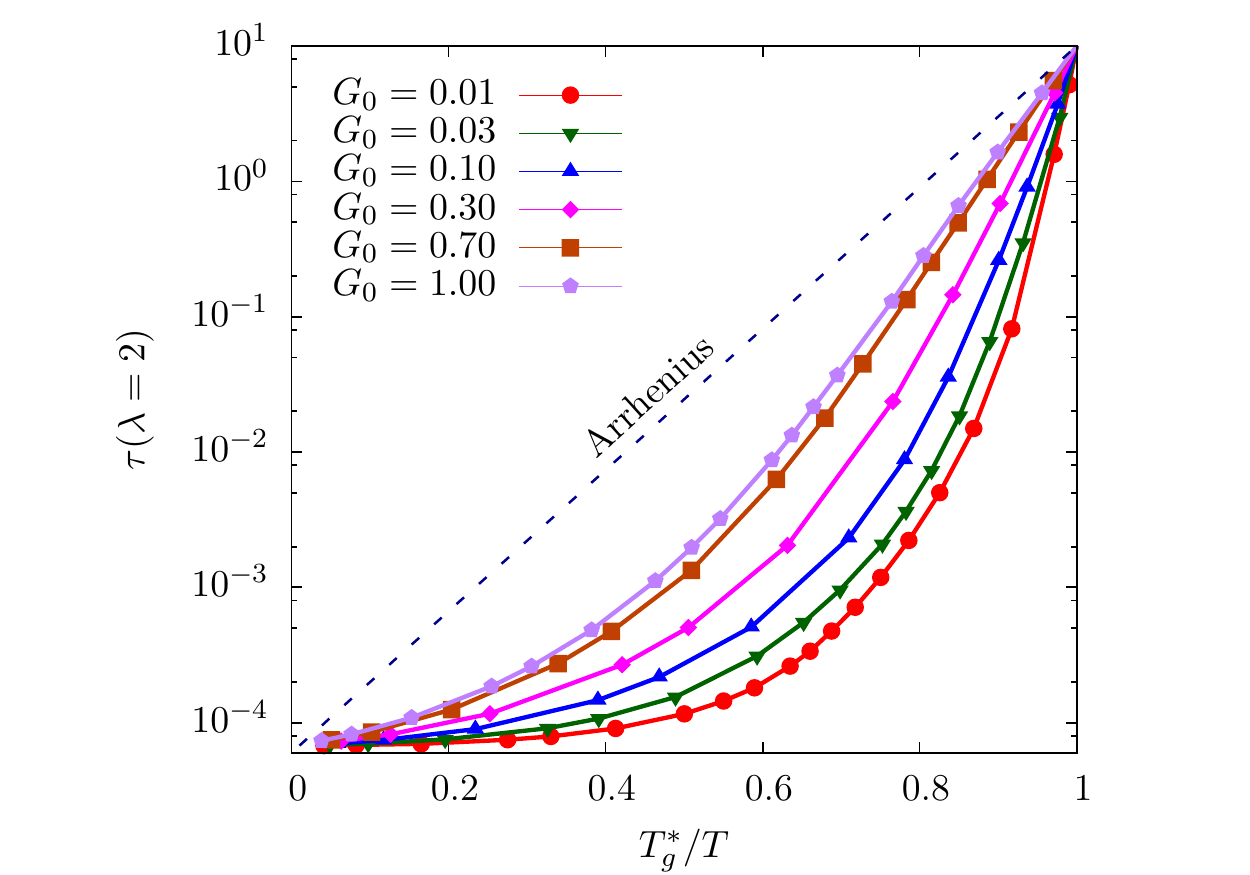}
	\caption{
		Angell plot of relaxation time $\tau$ for various $G_0$, with $\lambda = 2$ and $T^{*}_g$ defined at $\tau = 10$. 
	}
	\label{fig:Angelltau}
\end{figure}

We now express the thermodynamic properties of the DPLM based on these properties of the individual components. 
Using \eqr{N_A}{V_B} and  \eq{X}, \eq{Eeq} can be recast into:
\begin{equation}
\varepsilon_{eq} = \frac{z}{2}\squarebk{X\overline{V}_{A} + (1 - X)\overline{V}_{B}}.
\label{Eeq3}
\end{equation}
By differentiating it with respect to $T$, we arrive at
\begin{equation}
c_{v,eq} = \frac{z}{2}\squarebk{-\frac{dX}{dT}\roundbk{\overline{V}_{B} - \overline{V}_{A}} + X\frac{d\overline{V}_{A}}{dT}}.
\label{cveq3}
\end{equation}
after noting that ${d\overline{V}_{B}}/{dT}=0$

For $k_B T \ll \Delta V$, a condition under which the peak value of $c_{v,eq}$ occurs, \eq{V_A} reduces to
${\overline{V}_{A} \simeq V_0 + k_BT}$ and \eq{cveq3} becomes
\begin{equation}
c_{v,eq} \simeq \frac{z}{2}\squarebk{ -\frac{dX}{dT}\roundbk{\Delta V - k_BT} + Xk_B}.
\label{cveq4}
\end{equation}
Since $k_B=\Delta V = 1$ and $z=4$, the large peak value of $c_{v,eq} \simeq 18$ for $G_0=0.01$ in \fig{fig:cveqT} in fact results from a large magnitude of $dX/dT$.  Specifically, $| dX/dT |$ rises to 10 at $T \simeq 0.136$ corresponding to the peak of $c_{v,eq}$ as shown in \fig{fig:dXdT}.
This quantitatively demonstrates our suggestion that a large $c_{v,eq}$ for fragile glass is caused by a dramatic shift in the probabilistic weights of the two components in $g(V)$.


\section{Glassy dynamics}
\label{appendix:glassy_dynamics}
%

Following the procedures given in \Ref{lee2020}, we perform equilibrium simulations at various $G_0$ and $T$.
Then, we compute the self-intermediate scattering function defined as
\begin{equation}
	F_{s}(\mathbf{k},t) = \anglebk{e^{i\mathbf{k}\cdot(\mathbf{r}_l(t) - \mathbf{r}_{l}(0))}}
	\label{Fs}
\end{equation}
where $\mathbf{r}_{l}$ is the position of particle $l$ and $k = 2\pi/\lambda$ with wavelength $\lambda = 2$.
\Fig{fig:SIF} shows $F_{s}(\mathbf{k}, t)$ versus time $t$ for the fragile ($G_0 = 0.01$) and strong ($G_0 = 1$) glasses for various $T$.
The structural relaxation time $\tau$ is defined by ${F_{s}(\mathbf{k}, \tau) = e^{-1}}$.
One can define a relaxation-time-based glass transition temperature $T^{*}_g$ as the temperature at which $\tau$ reaches $\tau_r$, where $\tau_r=10$ is a long relaxation time taken as a  reference value.
\Fig{fig:Angelltau} shows an Angell plot of $\tau$ against $T^{*}_g/T$.
As seen, the super-Arrhenius nature, and thus also the fragility, are enhanced monotonically as $G_0$ decreases.
We observe that $G_0 = 0.01$ gives a fragile glass, while $G_0 = 1$ gives a moderately strong glass. 
Glass with more Arrhenius behaviors can be simulated by adding a non-zero positive energy barrier offset to  the particle hopping rate, as discussed in \cite{lee2020}, or by modifying  $g(V)$ appropriately, which will be reported elsewhere.

\begin{figure}[t]
	\includegraphics[width=\linewidth]{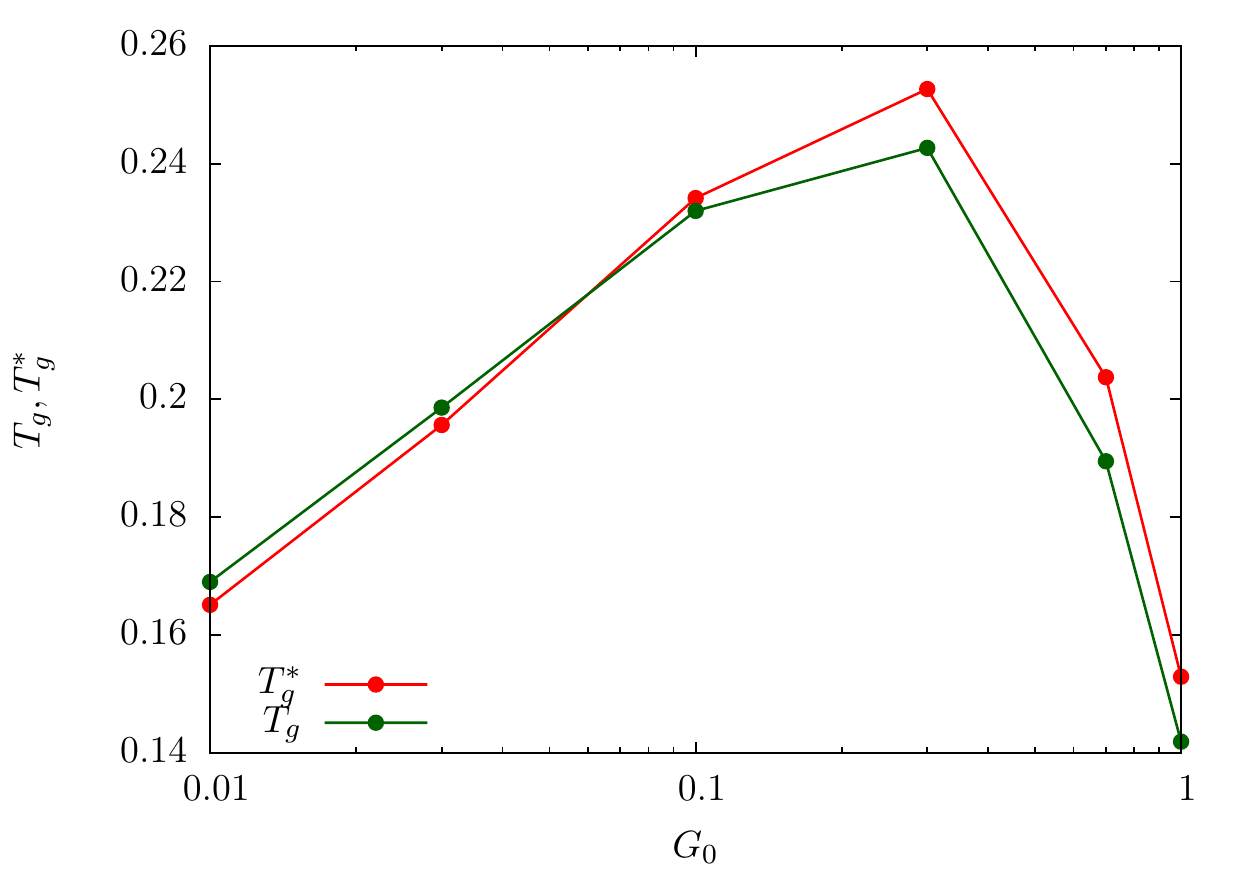}
	\caption{
		Comparison between heat-capacity-based $T_{g}$  measured at heating rate $\nu = 3\times 10^{-4}$ and relaxation-time-based $T^{*}_{g}$ with reference time $\tau_r=10$ for various $G_0$.
	}
	\label{fig:Tg_comparison}
\end{figure}

\Fig{fig:Tg_comparison} shows a comparison between the heat-capacity-based $T_{g}$ measured at $\nu = 3\times 10^{-4}$ and the relaxation-time-based $T^{*}_{g}$ with $\tau_r=10$ for various $G_0$.
As seen, $T_g$ and $T^*_g$ are quantitatively close to each other, as both $\nu = 3\times 10^{-4}$ and $\tau_r=10$ lead to similar modeled time scales, both of which leads to about  the longest simulations we can perform. Increasing $\nu$ or decreasing $\tau_r$ can lead to increases in $T_g$ and $T^{*}_{g}$ respectively.
On the other hand, the non-monotonic dependence of $T_g$ on $G_0$ has been explained in \Ref{lee2020}.
The Angell plot based on relaxation time $\tau$ in \fig{fig:Angelltau} gives a good indication of the kinetic fragility. Similar results are also obtained from a related Angell plot based on the diffusion coefficient $D$. 
Defining the glass transition as the point at which $D$ decreases to a reference value $D_r=0.1$, corresponding to the longest time scale we can simulate, the kinetic fragility $m_k$ for $G_0=0.01$ and 1 have been evaluated to be $26$ and 7 respectively  \cite{lee2020}. They should be compared with $m_k=4.7$ for an Arrhenius behavior under this definition.
To see what materials these models correspond to, results have been extrapolated to  a more realistic reference value of $D_r=10^{-14}$.
This gives $m_k \simeq 116$ for $G_0 = 0.01$ \cite{lee2020}, which is fragile and it is $4.5$ times larger than the unextrapolated value of 26.
The extrapolation scheme however cannot discriminate between the moderate strong glass at $G_0=1$ from a strong glass.
Instead, by analogy to the fragile glass, we simply estimate its fragility to be 4.5 times of the unextrapolated value, giving $m_k \simeq 31$.  We thus suggest that the glasses with $G_0=0.01$ and 1 model fragile and moderately strong glasses of fragilities around 116 and 31 respectively. 
Examples of them can be toluene and typical metallic glasses.

%

\end{document}